\documentclass[aps,10pt,superscriptaddress]{revtex4}%

\usepackage{amsfonts}
\usepackage{amsmath}
\usepackage{amssymb}
\usepackage{graphicx}%
\usepackage{mathptmx}
\def\bq{\begin{equation}}
\def\eq{\end{equation}}
\def\bqy{\begin{eqnarray}}
\def\eqy{\end{eqnarray}}

\def\deq{\ \dot=\ }

\def\de{\delta}

\def\ka{\kappa}

\def\Om{\Omega}
\def\p{\partial}

\def\p{\partial}

\def\bfb{\mathbf{B}}

\def\bfe{\mathbf{E}}

\def\bfj{\mathbf{J}}

\def\bfm{\mathbf{M}}

\def\bfx{\mathbf{x}}
\def\bfv{\mathbf{v}}

\def\bfs{\mathbf{s}}

\def\calc{\mathcal{C}}

\def\cale{\mathcal{E}}
\def\calf{\mathcal{F}}

\def\calk{\mathcal{K}}

  \def\Brac#1#2{\{#1,#2\}}
  \def\brac#1#2{[#1,#2]}
  \def\nest#1#2#3{\{\{#1,#2\},#3\}}

\begin{document}

\title{Gauge-Free Hamiltonian Structure of the Spin Maxwell-Vlasov  Equations}

\author{M. Marklund} 
\email[E-mail address: ]{mattias.marklund@physics.umu.se}
\affiliation{Department of Physics, Ume{\aa} University, SE--901 87 Ume{\aa}, Sweden}

\author{P. J. Morrison}
\email[E-mail address: ]{morrison@physics.utexas.edu}
\affiliation{Department of Physics and Institute for Fusion Studies, University of Texas, Austin, TX 78712, USA}

\date{\today}

\begin{abstract}
	We derive the gauge-free Hamiltonian structure of an extended kinetic theory, for which the intrinsic spin of the particles is taken into account. Such a semi-classical theory can be of interest for describing, e.g., strongly magnetized plasma systems. We find that it is possible to construct generalized brackets for the extended phase space, and discuss the implications of our findings.
\end{abstract}

\maketitle

High energy density plasma physics has become a popular subject (see, e.g., \cite{drake} and references therein). In such systems, quantum mechanical effects, such as wave function dispersion and/or statistical effects, can become important (for a recent experimental example, see \cite{glenzer-etal}), and much of these plasmas can be rightly termed \textit{quantum plasmas}. Much of the early literature on quantum plasmas has focused on condensed matter systems with a background lattice structure and the linear effects that follows (see, e.g., \cite{pines}). However, recent developments shows a different direction, where the nonlinear aspects of such systems are in focus \cite{manfredi,shukla-eliasson}. Examples of recent results include quantum ion-accoustic waves \cite{bains}, Jeans instabilities in quantum plasmas \cite{ren}, trapping effects \cite{shah}, magnetization by photons \cite{shukla} and relativistic effects \cite{melrose,masood}. Typically, the quantum hydrodynamic equations are derived by starting from the Schr\"odinger equation and making the Madelung ansatz \cite{manfredi-howto}. However, a method that more closely resembles the classical case is to use kinetic equations as a starting point (see Ref.\ \cite{manfredi-howto} for a comparison between the different approaches). 
The field of quantum kinetic theory \cite{carruthers} in many ways started with the ambitions of Wigner, as presented in Ref.\ [\onlinecite{wigner32}], to bridge the gap between classical Liouville theory and statistical quantum dynamics \cite{weyl,groenewold,moyal}. Thus, the development of quantum kinetic theory was partly due to an interest in obtaining a better understanding of the quantum-to-classical transition \cite{zachos}. However, another important aspect of quantum kinetic theory is as a computational tool for, e.g., quantum plasmas \cite{shukla-eliasson}, condensed matter systems \cite{Markowich,haug-jauho}, and, in general, quantum systems out of equilibrium \cite{rammer}, and in that respect shares many commonalities with quantum optics \cite{leonhardt}.
As shown in  \cite{cow, cow2, markbrod,brodmark,brodmark2}, spin is such an effect, the one of  particular interest here.
 
 When developing new models it can be important to show that they are Hamiltonian, since all of the most important models of physics have this property when phenomenological or other dissipation is neglected.  Matter models in terms of Eulerian  or spatial variables possess noncanonical Hamiltonian form, i.e.\  they are Hamiltonian but the conventional variables are not  a canonically conjugate set and consequently the Poisson bracket possesses  noncanonical form yet retains its Lie algebraic properties of antisymmetry, bilinearity, and the Jacobi identity. (See e.g.\ \cite{morgreene, morrison82, morrison98, morrison05}. Also see \cite{tassi2,tommaso}  for recent work applicable to plasmas.)  
 
 For the kinetic theory of interest here we show it has a Hamiltonian structure that is a generalization of that given in \cite{morrison80,morrison82,MW}.  (See also \cite{ibb2}.)\ \  We present the noncanonical Poisson bracket, prove directly that it satisfies the Jacobi identity, find Casimir invariants for the theory, and present an energy-like theorem that demonstrates that all equilibria with monotonically decreasing  distributions are stable.

\bigskip

We consider the nonrelativistic spin Maxwell-Vlasov equation for  $f(\bfx,\bfv,\bfs,t)$, an electron phase space density:
\bq
\frac{\p f}{\p t} = -\bfv \cdot\nabla f 
+ \left[ \frac{e}{m}\left(\bfe +\frac{\bfv}{c}\times\bfb\right) +\frac{2\mu_e}{m \hbar c} \nabla(\bfs\cdot\bfb)\right]\cdot\frac{\p f}{\p \bfv} +\frac{2\mu_e}{\hbar c}\left(\bfs\times\bfb\right)\cdot \frac{\p f}{\p \bfs}\
\label{smv}
\eq
where $m$ and $e>0$ are  the electron mass and charge, respectively, $2\pi\hbar$ is Planck's constant,   $\mu_e=g\mu_{\bfb}/2$  is the electron magnetic moment in terms of   $\mu_{\bfb}$,  the Bohr magneton,   and the electron spin $g$-factor.   Equation (\ref{smv}) is coupled to the dynamical Maxwell equations, 
\bqy
\frac{\p \bfb}{\p t}&=&-\nabla\times \bfe
\label{farad}
\\
\frac{\p \bfe}{\p t}&=&\nabla\times \bfb-4\pi \bfj
\label{amp}
\eqy
 through  the current $\bfj=\bfj_f+\nabla \times \bfm$, which  has ``free'' and spin magnetization parts:
\bqy
\bfj_f&:=&-e\int\!\!d^3v\,d^3s \,\, \bfv  f
\label{jf}
\\
\bfm&:=&-\frac{2\mu_e}{\hbar c} \int\!\!d^3v\,d^3s \,\,\bfs f \,.
\label{mag}
\eqy
Extension to multiple species is straightforward. 

Note,  Eqs.~(\ref{smv}), (\ref{farad}), and (\ref{amp}), with (\ref{jf}) and (\ref{mag}),  are to be viewed classically and consequently a full nine-dimensional phase space integration,  $d^9z=d^3xd^3vd^3s$,  is considered for $f$.   Later we will see how a spin quantization constraint can be applied. 

The Hamiltonian functional for the theory is
\bq
H[\bfe,\bfb,f]=\int\!\!d^9z\, \left(\frac{m}{2} v^2 + \frac{2\mu_e}{\hbar c} \bfs \cdot \bfb\right) f + \frac1{8\pi}\int\!\! d^3x\, \left(E^2 + B^2\right)
\eq
which can be shown directly to be conserved, but this will become obvious after the Hamiltonian structure is given.

The noncanonical spin Maxwell-Vlasov bracket is composed of several parts:
\bqy
\Brac FG_{sMV}&=&\int \! d^9z  \, f \Big(\brac{F_f}{G_f}_c 
\label{cano}\\
&{\ }& \hspace{ .6 in} +  \brac{F_f}{G_f}_{B} 
\label{mw}\\
&{\ }& \hspace{ 1.1 in} +  \brac{F_f}{G_f}_s
\label{new}   \\
&{\ }& \hspace{ 1.4 in}  + \frac{4\pi e}{m}\left(F_E\cdot\p_v G_f - G_{E}\cdot\p_v F_f\right) \Big)
\label{couple}\\
&{\ }& \hspace{ 1.75 in} +4\pi \!\int\!d^3x \, \left(F_E\cdot \nabla\times G_B - G_E\cdot\nabla\times F_B\right), 
 \label{MVSbkt}
\eqy
where  
\bqy
\brac fg_c&:=&\frac1{m}\left(\nabla f\cdot\p_vg - \nabla g\cdot\p_vf\right)\,,
\label{canbkt}
\\  
\brac fg_B&:=&-\frac{eB}{m^2c}\cdot\left(\p_v f\times\p_v g\right)\,,
\label{bbkt}
\\
\brac fg_s&:=&    s\cdot (\p_s f\times \p_s g)\,, 
\label{pspin}
\eqy
with standard partial derivatives   denoted  by $\p_{v}:={\p}/{\p \bfv}$  and functional derivatives by $F_f:={\de F}/{\de f}$,   etc. Term (\ref{new}) of $\{\,,\,\}_{sMV}$ is new and accommodates the spin; it is not surprising that it has an inner bracket based on the $so(3)$ algebra (\cite{sudarshan}).  The remaining terms (\ref{cano}),  (\ref{mw}), (\ref{couple}),  and (\ref{MVSbkt})   produce the usual Vlasov-Maxwell theory \cite{bornin,morrison80,morrison82,MW,ibb2}. 
 It is a simple exercise to show that Eqs.~(\ref{smv}), (\ref{farad}), and (\ref{amp}) are given as follows:
 \bqy
 \frac{\p f}{\p t}&=&\{f,H\}_{sMV}
 \nonumber\\
 \frac{\p \bfb}{\p t}&=&\{\bfb,H\}_{sMV}
  \nonumber\\
 \frac{\p \bfe}{\p t}&=&\{\bfe,H\}_{sMV}\,.
  \nonumber
  \eqy
This is facilitated by the identity $\int d^9z\,  f[g,h]=-\int d^9z\, g[f,h]$,  which works for all three brackets  of  (\ref{canbkt}),   (\ref{bbkt}),  and (\ref{pspin}).

 There are two approaches to obtaining a Hamiltonian description.  The usual way is by constructing an action principle by postulating a Lagrangian density with the desired observables and symmetery group, and then effecting a Legendre transformation, when possible, to obtain a Hamiltonian theory.  Alternatively one can postulate an energy functional and Poisson bracket as we have done here.  When exploring new territory with this latter approach, one must prove directly the Jacobi identity $\nest FGH + \nest GHF+ \nest HFG \equiv 0$ for all functionals $F$, $G$, and $H$.  With the former approach this is guaranteed if the action principle and Legendre transform exist and one can perform a chain rule calculation to obtain a bracket in terms of the desired observables.    This was done for the Maxwell-Vlasov bracket in \cite{MW},  where it is necessary to assume the existence of a vector potential.  However, with the bracket approach one need not assum the existence of a vector potential, in which case the Maxwell-Vlasov bracket satisfies   
 \bq
\{\{F,G\}_{MV},H\}_{MV} + {\rm cyc}= \int\!d^6z \, f\, 
\nabla\cdot \mathbf{B} \left[ \left(\frac{\p F_f}{\p \mathbf{v}}\times \frac{\p G_f}{\p \mathbf{v}}\right)\cdot \frac{\p H_f}{\p \mathbf{v}}\right]\,.
\label{mvjac}
\eq
This result was quoted in   \cite{morrison82} -- see \cite{morrison10} for a recent recounting of the explicit (and tedious) details of this early calculation.   Thus, although the Maxwell-Vlasov Hamiltonian theory is gauge-free, it requires $\nabla \cdot \mathbf{B}=0$.  

One can construct an action principle for the spin Maxwell-Vlasov theory of the form of \cite{P84,PM85,PM91} and then proceed  to the bracket $\{F,G\}_{sMV}$ (see e.g.\ \cite{YeM,morrison10}), but we find it easier to prove  the Jacobi identity directly. Writing $\{F,G\}_{sMV}=\{F,G\}_{MV} + \{F,G\}_{s}$ and using $:=:$ to denote the cyclic sum we have
\bqy
\{\{F,G\}_{sMV},H\}_{sMV}&:=:& \{\{F,G\}_{MV},H\}_{MV} + \{\{F,G\}_{s},H\}_{MV} +
\nonumber\\
&{\ }& \qquad+ \{\{F,G\}_{MV},H\}_{s} + \{\{F,G\}_{s},H\}_{s}
\nonumber\\
&:=:& \{\{F,G\}_{s},H\}_{MV} 
+ \{\{F,G\}_{MV},H\}_{s}\,,
\eqy
where the second equality follows because of (\ref{mvjac}) (assuming solenoidal $\mathbf{B}$) and the fact that $\{F,G\}_{s}$ is a Lie-Poisson bracket (see e.g.\ \cite{morrison98,marsden}).  Thus it only remains to show that the cross terms cancel, which is facilitated by a theorem in \cite{morrison82}; viz., when functionally differentiating $\{F,G\}_{MV}$ and $\{F,G\}_{s}$, which are needed when constructing the cross terms,  one can ignore the second functional derivative terms.  These cancel by virtue of the symmetry of the second variation and antisymmetry of the bracket.  Using the symbol $\deq$ to denote equivalence modulo the second variation terms, we obtain
\bqy
\frac{\de \{F,G\}_{MV}}{\de f}&\deq& [F_f,G_f]_c + [F_f,G_f]_B
+ \frac{4\pi e}{m}\left(F_E\cdot\p_v G_f - G_{E}\cdot\p_v F_f\right)
\\
\frac{\de \{F,G\}_{s}}{\de f}&\deq& [F_f,G_f]_s  \,,
\eqy
while all other needed functional derivatives vanish.  Thus
\bqy
\{\{F,G\}_{MV},H\}_{s}&:=:& \int\!\!d^9z\,\Big(f \, \big[[F_f,G_f]_c+ [F_f,G_f]_B, H_f\big]_s
\nonumber\\
&{\ }& \hspace{1.0 in} +  \frac{4\pi e}{m} f\big[F_E\cdot\p_v G_f - G_{E}\cdot\p_v F_f,H_f\big]_s\Big)
\label{jac1}\\
\{\{F,G\}_{s},H\}_{MV}&:=:& \int\!\!d^9z\,  \Big( f \big[[F_f,G_f]_s, H_f\big]_c +  f \big[[F_f,G_f]_s, H_f\big]_B\Big)
\nonumber\\
&{\ }& \hspace{1.0 in} +   \frac{4\pi e}{m} f \,  H_E\cdot\p_v [F_f,G_f]_s\Big)]\,.
\label{jac2}
\eqy
The first lines of (\ref{jac1}) and (\ref{jac2}) cancel by virtue of the Jacobi identities for the brackets $[\, ,\, ]_{c,B,s}$ on functions, while the second line of (\ref{jac1}) cancels upon permutation of  the second term. Similarly, the second term of (\ref{jac2}) vanishes. 

  Having established the Jacobi identity,  we search for Casimir invariants, functionals that commute with all other functionals. Using the equations obtained from $\{C, F\}= 0$ for all $F$,  we obtain
\bqy
   C^{fs}&=&\int d^9z\,  \calc(f,s^2)\,,
\\
   C^{E}&=&\int d^3x\, \kappa_E(\bfx)\left(\nabla\cdot \bfe + 4\pi e\int\!\!d^3v\, d^3s\, f \right)\,,
   \\
  C^{B}&=&\int d^3x\, \kappa_B(\bfx) \, \nabla\cdot \bfb\,,
   \eqy
where $\calc$, $\ka_E$, and $\ka_B$ are arbitrary functions of their arguments.   The Casimir $C^{fs}$ is a consequence of the fact that the solution to (\ref{smv}) is a volume preserving rearrangement, i.e.\ that the solution can be written as the initial condition on the characteristics.  It is not difficult to see that (\ref{smv}) can be written in conservation form on the full nine-dimensional space.  The $s^2$ dependence of the Casimir $C^{fs}$ is the lift of the $so(3)$ spin Casimir to the kinetic theory. Such inner Casimirs are always give rise to Casimirs of the field theory. The Casimir $C^E$ is of course Poisson's equation, an initial condition that would remain preserved should we change the Hamiltonian functional.  It is  a local Casimir because of the arbitrary function $\kappa_E(\bfx)$, which is used here to make the point that it is conserved point-wise.  The local quantity $C^{B}$ is technically not the same as the others because its vanishing is required for the Jacobi identity. However, this is only technical because $\{C^{B}, F\}=0$,  for all $F$,  whether or not  $\nabla\cdot \mathbf{B}=0$.

A consequence of the  Casimir $C^{fs}$ is  that $s^2$ is constant on level sets (contours) of $f$, which can be viewed as a classical prequantization property.  If we suppose $f$ has the from $f=c(s^2)f_c(\bfx,\bfv,\bfs,t)
$, then it follows that if $f_c$ satisfies (\ref{smv}) then $f$ does.  Choosing  
\bq
f=\de(|\bfs| - \hbar/2)f_c(\bfx,\bfv,\bfs,t)
\eq
we enforce the usual quantization condition and our integrals reduce from integrations over $d^9z$ to $d^3x,d^3vd\Om$, where $d\Om$ denotes the spin sphere as in e.g.\  \cite{brodmark}. Because of the pure antisymmetry of the $so(3)$ structure constants, Liouville's theorem on characteristics follows immediately; however, for general cosymplectic forms, $J$, i.e.\ for brackets of the form $[f,g]=\p f /\p w^i J^{ij}(w) \p g  /\p w^j$, one can insert a factor of $\sqrt{\det{J}}$ restricted to symplectic leaves to define a proper `volume' measure (see e.g.\  \cite{morrison98}).

Having found the Casimir invariants we can write down a variational principle for equilibria and then proceed to investigate stability by the technique introduce in \cite{KO} (see also \cite{gardner63}), which has become known as the energy-Casimir method  (see e.g.\ \cite{holm, moreli,morrison98}).  First we seek extrema of the quantity  $\calf:=H + C^{fs}+ C^{E}+ C^{B}$, which must give rise to equations for equilibria:
\bqy
\frac{\de \calf}{\de f}&=& \calk +4\pi e \ka_E +\calc_f(f,s^2)=0
\\
\frac{\de \calf}{\de \bfe}&=& \bfe - 4\pi \nabla \ka_E  =0
\\
\frac{\de \calf}{\de \bfb}&=& \bfb - 4\pi \nabla ka_B =0\,,
\eqy
where $\calc_f:=\p \calc/\p f$, $\calk:=m v^2/2 +  {2\mu_e} \bfs \cdot \bfb/{\hbar c}$ and we define the `particle energy' by $\cale:=\calk +4\pi  \ka_E$.  Evidently $-4\pi\ka_E$ is  the electrostatic potential and $\bfb$ must be an extenal field, i.e.\ $\bfj=\nabla\times\bfb=0$ (cf.\ the results for the Maxwell-Vlasov case \cite{MP89,MP90}). Assuming $\calc_f$ has an inverse,  we obtain the following for the equilibrium distribution function:
\bq
f_e(\cale)= \calc_f^{-1}(-\cale,s^2)\,.
\eq
If we chose $\calc$ to be proportional to the usual entropy expression $f\ln f$, neglect the dependence on $s^2$, and  assume $\bfe=0$,  an acceptible choice,  then we obtain the Maxwell-Boltzmann-like equilibrium of \cite{brodmark}.  Proceeding to the second variation we obtain
\bqy
\de^2\calf&=& \frac1{2}\int d^9z\, \calc_{ff}(\de f)^2 +\frac1{8\pi}\int\!\!d^3x \left((\de E)^2 + (\de B)^2 \right)
\nonumber\\
&=&- \frac1{2}\int d^9z\, \frac{(\de f)^2}{\p f_e/\p \cale} +\frac1{8\pi}\int\!\!d^3x \left((\de E)^2 + (\de B)^2 \right)\,,
\label{KOS}
\eqy
where the second equality of (\ref{KOS}) follows upon differentiating the condition $\cale + \calc_f=0$ with respect to $f$. From (\ref{KOS}) we immediately draw the formal conclusion that equilibria that are monotonically decreasing functions of $\cale$ are stable, because $\de^2\calf$ serves as a Lyapunov functional.  More rigorous versions of this have been proved  for the Vlasov equation in both the plasma and astrophysical contexts (see e.g.\ \cite{rein}). 

Only a limited class of equilibria come from $\de \calf=0$, and the reason  for this is somewhat subtle.  A complete set of equilibria  can be gotten from a constrained variational principle with   `dynamically accessible variations' \cite{morrison98}, but this will not be considered further here.

In summary, the formulation of an extended kinetic theory for electrons, taking into account the intrinsic spin and the relevant magnetization effects, was considered. In particular, the semiclassical limit, valid for length scales large   compared to the size of the electron wave function was given. Based on the extended phase space, the Hamiltonian structure was discussed, and a noncanonical Poisson bracket was found that satisfies the Jacobi identity. Furthermore, we obtained the related Casimir invariants and showed the stability of all equilibria with monotonically decreasing distributions. Our findings could act as a guiding tool for further extended Hamiltonian theories, including quantum effects from Pauli or Dirac theory, for which a gauge-line has to be included in the phase of the definition of the corresponding Wigner function \cite{haas-etal,zamanian-etal}. Moreover, the stability of the equilibria can be an important principle in future numerical studies of strongly magnetized systems.

\section*{Acknowledgment}
\noindent  PJM was supported by U.S. Dept.\ of Energy Contract \# DE-FG05-80ET-53088. MM was supported by the Swedish Research Council Contract \# 2007-4422 and the European Research Council Contract \# 204059-QPQV

\bibliographystyle{plain}

\bibliography{spin}

\end{document}